\documentclass[11pt]{article}
\usepackage{amsmath,amssymb,color,slashed}
\usepackage{amsfonts}
\newcommand{\hoch}[1]{$\, ^{#1}$}

\newcommand{\be}{\begin{equation}}
\newcommand{\ee}{\end{equation}}
\newcommand{\bea}{\setlength\arraycolsep{2pt} \begin{eqnarray}}
\newcommand{\eea}{\end{eqnarray}}

\def\ft#1#2{{\textstyle{\frac{\scriptstyle #1}{\scriptstyle #2} } }}
\def\fft#1#2{{\frac{#1}{#2}}}

\def\0{{\sst{(0)}}}
\def\1{{\sst{(1)}}}
\def\2{{\sst{(2)}}}
\def\3{{\sst{(3)}}}
\def\4{{\sst{(4)}}}
\def\5{{\sst{(5)}}}
\def\6{{\sst{(6)}}}
\def\7{{\sst{(7)}}}
\def\8{{\sst{(8)}}}
\def\sst#1{{\scriptscriptstyle #1}}

\begin{document}

\begin{flushright}
\hfill{CAS-KITPC/ITP-285 \ \ \ \ KIAS-P11072}
 %\hfill{
%\bf hep-th/yymmnnn}
\end{flushright}

\vspace{25pt}
\begin{center}
{\large {\bf $f(R)$ Gravities, Killing Spinor Equations,\\ ``BPS''
Domain Walls and Cosmology }}

\vspace{10pt}

Haishan Liu\hoch{1}, H. L\"u\hoch{2,3} and Zhao-Long Wang\hoch{4}

\vspace{10pt}

\hoch{1}{\it Zheijiang Institute of Modern Physics\\
Department of Physics, Zhejiang University, Hangzhou 310027, China}

\vspace{10pt}

\hoch{2}{\it China Economics and Management Academy\\
Central University of Finance and Economics, Beijing 100081, China}

\vspace{10pt}

\hoch{3}{\it Institute for Advanced Study, Shenzhen University\\
Nanhai Ave 3688, Shenzhen 518060, China}

\vspace{10pt}

\hoch{4} {\it School of Physics, Korea Institute for Advanced Study,
Seoul 130-722, Korea}

\vspace{40pt}

\underline{ABSTRACT}
\end{center}

We derive the condition on $f(R)$ gravities that admit Killing
spinor equations and construct explicit such examples. The Killing
spinor equations can be used to reduce the fourth-order differential
equations of motion to the first order for both the domain wall and
FLRW cosmological solutions.  We obtain exact ``BPS'' domain walls
that describe the smooth Randall-Sundrum II, AdS wormholes and the
RG flow from IR to UV. We also obtain exact smooth cosmological
solutions that describe the evolution from an inflationary starting
point with a larger cosmological constant to an ever-expanding
universe with a smaller cosmological constant. In addition, We find
exact smooth solutions of pre-big bang models, bouncing or crunching
universes. An important feature is that the scalar curvature $R$ of
all these metrics is varying rather than a constant. Another
intriguing feature is that there are two different $f(R)$ gravities
that give rise to the same ``BPS'' solution. We also study
linearized $f(R)$ gravities in (A)dS vacua.

\vspace{15pt}

\thispagestyle{empty}

\pagebreak

\tableofcontents

\addtocontents{toc}{\protect\setcounter{tocdepth}{2}}

%%%%%%%%%%%%%%%%%%%%%%%%%%%%%%%%%%%%%%%%

\newpage
%%%%%%%%%%%%%%%%%%%%%%%%%%%%%%%%%%%%%%%%

\section{Introduction}

Modifying Einstein gravity with higher-order curvature invariants
goes back to early days of General Relativity, notably by Eddington
and Weyl simply as an exercise of intellectual curiosity
\cite{eddweyl}. A natural and perhaps the simplest generalization of
Einstein gravity is to replace the Ricci scalar $R$ in the
Einstein-Hilbert action with an arbitrary function of $R$
\cite{berg,rr,bgs,buch}. The resulting $f(R)$ theories of gravity
can admit Einstein metrics with a positive, negative or zero
cosmological constant.  This makes $f(R)$ gravities versatile in
both reproducing Newtonian gravity and studying cosmology, although
a convincing $f(R)$ theory that fits all the observational data
remains elusive. It can be shown \cite{ohan,teytou} that $f(R)$
gravity is equivalent to some special class of the Brans-Dicke
gravity/scalar theory \cite{bradic} by the Legendre transformation;
however, the conversion requires to find the inverse function of
$f'(R)$ which may not have a close analytical form. Thus in general
$f(R)$ gravity should be studied on its own right.  Since an
inflationary model involving the quadratic Ricci scalar was
constructed first time in \cite{star}, there has been ongoing
interest in $f(R)$ gravities for the last three decades, and the
application has been focused largely on the area of cosmology. (See,
{\it e.g.}, reviews \cite{rev0,rev1,rev2}.)

    The fact that AdS spacetimes arise naturally in $f(R)$ gravities
suggests that they can also be used for investigating the AdS/CFT
correspondence \cite{mald,gkp,wit}.  However, apart from those with
constant scalar curvature, exact solutions in $f(R)$ gravities are
difficult to come by. When the Maxwell field is introduced in the
theory, charged black holes in four dimensions were obtained
recently \cite{bh1,bh2,bh3}. This is possible only because the
Maxwell field in four dimensions does not contribute to the trace of
the Energy-momentum tensor so that the Ricci scalar $R$ in the
four-dimensional charged black holes remains constant.  It is easy
to see that if $f(R)$ gravity coupled to a matter system whose
energy-momentum tensor has vanishing trace, the theory admits
analogous solutions as those in Einstein gravity coupled to the
matter. Such solutions with constant $R$ can be viewed as somewhat
trivial since they do not explore the nature and properties of the
function $f$. Non-trivial solutions with varying $R$ have been
hitherto unknown in $f(R)$ gravities. With few interesting and
non-trivial exact solutions, the effort in applying the AdS/CFT
correspondence in $f(R)$ gravities has been severely limited.

    In supergravities, owing to the existence of Killing
spinor equations, much wider classes of exact BPS solutions have
been constructed. This is because the Killing spinor equations can
be loosely viewed as the first integral of Einstein's second-order
equations of motion, and hence they significantly simplify the
equations. Killing spinor equations are not exclusive for
supergravities. Einstein theory of gravity with or without a
cosmological constant in any dimension admits a Killing spinor
equation.\footnote{The existence of Killing spinor equations in
Einstein gravity or even in supergravity does not imply that all
background solutions have Killing spinors which are local
solutions of the Killing spinor equations. Furthermore, manifolds without spin structure would not admit spinors at all. The backgrounds that admit Killing spinors are called BPS solutions in supergravities, whilst manifolds with Killing spinors in Einstein theory are referred to as ones with the reduced holonomy.  The criteria on a theory that admits Killing
spinor equations will be discussed in section 2.3.} Killing spinor
equations were also known to exist in the gravity/scalar system
where the scalar potential is constructed in terms of a
superpotential \cite{fnss}. It has been recently demonstrated that
even when involving form fields, some non-supersymmetric gravity
theories can admit Killing spinor equations \cite{lpw,lw,llw}.
Examples are rare, but the low-energy effective action of the
bosonic string up to the $\alpha'$ order was shown to admit Killing
spinor equations in any dimension. These Killing spinor equations
allow one to find new classes of solutions in these theories.

   Obviously, a generic $f(R)$ theory does not admit Killing spinor
equations.  On the other hand, there must exist subclasses of $f(R)$
gravities that do. The simplest example is the aforementioned
Einstein gravity with/out a cosmological constant.  In this paper,
we follow the technique developed in \cite{lpw,lw,llw} and propose
the two Killing spinor equations for $f(R)$ gravities. (Our paper
deals only with the $f(R)$ theories in the metric formalism.) They
involve two functions $W$ and $U$ of $R$.  We then derive the
condition on $f(R)$ so that the theory admits these equations.  We
find that for a given choice of $W$, the function $f$ satisfies a
second-order linear differential equation, implying that there exist
one (non-trivial) parameter family of $f(R)$ gravities for the same
$W$.  The function $U$ is then fully determined by $W$ and $f$.
Although there are no analytical solutions to the second-order
differential equation in general, we find many explicit examples of
$f(R)$ gravities that do admit Killing spinor equations.

The advantage of having Killing spinor equations in our $f(R)$
gravities is that they can be used to construct a large class of
solutions involving only first-order differential equations.  These
solutions are analogous to the BPS solutions in supergravities, and
hence we refer them as ``BPS'' even though our $f(R)$ gravities are
not supersymmetric.  The focus of our construction is the static
domain wall and the FLRW (Friedmann-Lema\^itre-Robertson-Walker)
cosmological solutions, both of which are conformally flat and of
cohomogeneity one. It turns out that these ``BPS'' solutions are
solely determined by the function $W$ in the Killing spinor
equations. It follows that there are two different $f(R)$ gravities
that give rise to the same solution. We use explicit examples to
compare the pros and cons of such two $f(R)$ theories.  Another
characteristic of our $f(R)$ gravities is that they admit at least
two (A)dS vacua, and the solutions we construct are typically
smooth, running from one (A)dS vacuum to the other. They provide
excellent examples for studying either the AdS/CFT correspondence or
cosmology. It should be pointed out that although multiple (A)dS
vacua exist also in Lovelock gravities with the Gauss-Bonnet type of
topological terms, there is no known flow that links these vacua.
This is one major difference between our $f(R)$ gravities and the
Lovelock type of theories.

This paper is organized as follows.  In section 2, we first give a
quick review of $f(R)$ gravities and the equations of motion. We
then demonstrate a phenomenon with an explicit example that there
can be new types of (A)dS vacua with vanishing $f(R_0)$ but
divergent $f'(R_0)$. The same vacuum can be embedded in a different
theory with no such singular behavior. We then propose Killing
spinor equations and derive the $\Gamma$-matrix projected
integrability conditions.  These allow us to derive the condition on
$f$ that admits Killing spinor equations.  We then give a few
examples of such theories.  More examples will be given in
subsequent sections. In section 3, we consider ``BPS'' domain wall
solutions. We obtain a class of exact solutions that connect two AdS
vacua. These solutions include the smooth Randall Sundrum II, AdS
wormholes and the RG flow from the IR (infrared) region to the UV
(ultraviolet) region. These explicit solutions demonstrate that
$f(R)$ gravities are quite attuned to the investigation of the
AdS/CFT correspondence.

       In section 4, we examine the FLRW solution with
flat spatial directions.  We find that Killing spinor equations can
also be useful to simplify the time-dependent equations. We obtain
large classes of exact cosmological solutions. One type describes
the evolution from an inflationary starting point to end with an
ever-lasting expanding universe, very much like our Universe. We
also find smooth metrics of pre-big bang models, bouncing and
crunching universes.

       In section 5, we give a couple of examples of converting our
$f(R)$ gravities to the Brans-Dicke theory.  In these examples, the
inverse function of $f'(R)$ can be obtained as simple analytical
functions. The majority of our $f(R)$ gravities we obtained in this
paper do not give rise to a close form scalar potential in the
corresponding Brans-Dicke theory. In section 6, we study the linear
spectrum of $f(R)$ gravities in (A)dS vacua.  As one would expect,
in general the spectrum consists of the massless spin-2 graviton and
a massive trace scalar mode.  We derive the ghost-free and
tachyon-free conditions. We discuss the special circumstance where
the spectrum becomes less straightforward with the kinetic terms for
the graviton and/or the scalar modes dropped from the linearized
action. One intriguing feature of our construction is that there
typically exist two $f(R)$ gravities for a given ``BPS'' solution.
For such a solution that connects to two different (A)dS vacua, each
of the two theories is suitable for one of the two vacua
respectively. We conclude the paper in section 7.

\section{$f(R)$ gravities with Killing spinor equations}

\subsection{Lagrangian and equations of motion}

The Lagrangian for $f(R)$ gravity in $D$ dimensions is
\begin{equation}
{\cal L}_D=\sqrt{-g} f(R)\,,
\end{equation}
where $f$ is a generic real function.  In this paper, we shall be
concerned with only $f(R)$ theories in the metric formalism, and hence
the equations of motion from the variation of $g_{\mu\nu}$ are given by
\begin{equation}
{\cal G}_{\mu\nu}\equiv F(R) R_{\mu\nu} - \ft12 f(R) g_{\mu\nu} +
(g_{\mu\nu}\Box - \nabla_\mu\nabla_\nu) F(R)=0\,,\label{Gmunueom}
\end{equation}
where $F(R)=f'(R)$.  Note that in this paper, we always use a prime
to denote a derivative with respect to $R$, unless an explicit new
variable is given. Taking the trace, we have
\begin{equation}
{\cal R}\equiv RF - \ft12 D f + (D-1) \Box F=0\,.\label{traceeom}
\end{equation}
The equations of motion (\ref{Gmunueom}) can be equivalently
expressed as
\begin{equation}
{\cal R}_{\mu\nu}\equiv R_{\mu\nu} - \fft{1}{F} \nabla_\mu\nabla_\nu
F+ \fft{1}{2(D-1)F} ( f-2R F)g_{\mu\nu}=0\,.\label{Rmunueom}
\end{equation}
Note that although $f(R)$ gravity can be related to the Brans-Dicke
theory by the Legendre transformation, the resulting scalar
potential involves the inverse function of $F$, which may not have a
close analytical expression. Thus for general $f(R)$ gravities, the
theories are best studied in their original forms rather than
converting them to the corresponding unnatural gravity/scalar
system.  We shall come back to this point in section 5.

\subsection{New (A)dS vacua in $f(R)$ gravities}

The simplest class of solutions for $f(R)$ gravities are perhaps the
metrics with constant $R$, which we denote as $R_0$.  In general,
the metrics are Einstein, {\it i.e.} $R_{\mu\nu}=\Lambda
g_{\mu\nu}$, with the effective cosmological constant
$\Lambda=R_0/D$. It follows from (\ref{traceeom}) that
\begin{equation}
2R_0 F(R_0)=Df(R_0)\,.\label{R0eom}
\end{equation}
Depending on whether $\Lambda$ is positive, 0, or negative, the
vacuum solution is de Sitter (dS), Minkowski and anti-de Sitter
(AdS) respectively.  In the special case when both $f(R_0)$ and
$F(R_0)$ vanish, the equations of motion (\ref{Gmunueom}) are
satisfied simply by $R=R_0$. This degenerate case allows any metric
with constant scalar curvature $R_0$ to be a solution, including
some Lifshitz black holes \cite{lif1,lif2}.  As we shall see in
section 6, such an (A)dS vacuum has no propagating spin-2 graviton
mode, but only a scalar trace mode. (See \cite{nood} for a review on
solutions in $f(R)$ theories.)

     In this subsection, we demonstrate that new classes of (A)dS
solutions can emerge in $f(R)$ gravity, which are not solutions of
(\ref{R0eom}), but characterized by the divergent $F(R_0)$.  To
illustrate this, let us consider the following $f(R)$ theory
\begin{equation}
{\cal L}_4= \sigma_1 \sqrt{-g}\, R \sqrt{48\beta^2 -
R}\,,\label{d4prelag}
\end{equation}
It follows from (\ref{R0eom}) that the theory has two vacua with
\begin{equation}
R_0=0\qquad\hbox{and}\qquad R_0=96\beta^2\,.\label{twoR0}
\end{equation}
It is clear that the two vacua cannot be connected, since for a
given $\sigma_1$, the Lagrangian can only be real either at the
vicinity of $R_0=0$ or $R_0=96\beta^2$, but not at both. As we shall
see later, this theory admits Killing spinor equations which enable
us to find an exact cosmological solution with varying $R$:
\begin{eqnarray}
ds^2_4 &=& - dt^2 + a^2  (dx_1^2 + dx_2^2 + dx_3^2)\,,\cr %%
a^2 &=& 1 + e^{4\beta(t-t_0)}\,.\label{d4premetric}
\end{eqnarray}
It is a straightforward exercise to verify that the metric satisfies
(\ref{Gmunueom}).  Let us consider the case with $\beta >0$.  The
metric is Minkowski when $t\rightarrow -\infty$, and it runs to the
de Sitter spacetime with $R_0=48\beta^2$ when $t\rightarrow
+\infty$.  The Ricci scalar increases monotonically with respect to
$t$. The derivation of this solution can be found in section 4.4.

  The solution (\ref{d4premetric}) describes a pre-big bang model
without singularity. Long before the inflation starts, the universe
is Minkowski under the perturbative Lagrangian of Einstein gravity
with higher-order Ricci curvature terms
\begin{equation}
{\cal L}_4=4\sqrt{3}\,\beta\sigma \sqrt{-g} \Big(R
-\fft{R^2}{96\beta^2} + \cdots\Big)\,.
\end{equation}
The universe bursts into inflation around $t=t_0$ when the
non-perturbative effect takes place.

What we would like to draw attention to here is that if we rescale
$x_i\rightarrow e^{2\beta t_0} x_i$ and then send the integration
constant $t_0\rightarrow -\infty$, we arrive at a dS metric
\begin{equation}
ds^2 = -dt^2 + e^{4\beta t} (dx_1^2 + dx_2^2 + dx_3^2)\,,\label{d4ads}
\end{equation}
with $R_0=48\beta^2$. It is clear that this is a limiting solution
of our $f(R)$ gravity; however, it is not a solution of
(\ref{R0eom}), which gives only (\ref{twoR0}).  This solution is
characterized by $f(R_0)=0$ and $F(R_0)=\infty$, and hence the
equation (\ref{traceeom}) breaks down when $R\rightarrow R_0$.

In this paper, we are able to uncover such (A)dS solutions because
our Killing spinor equations allow us to find exact solutions with
non-constant $R$. New (A)dS metrics emerge when we let $R$ run to
such $R_0$.  It is not obvious to us how to find such a solution in
the situation when we cannot obtain an exact solution with running
$R$.  It is a subject worth further investigation.  It should be
pointed out that from the point of view of the Brans-Dicke theory,
the solution should be considered as singular since the $F(R)$
corresponds to the scalar mode and it blows up at $R=R_0$.  One can
also take a different point of view that such a theory is
intrinsically pure gravity.

Interestingly, we find that there is another quite different $f(R)$
theory that can give rises to exactly the same cosmological solution
(\ref{d4premetric}), namely
\begin{equation}
{\cal L}_4 = \sigma_2 \sqrt{-g}\, R\Big[12\beta -
\sqrt{3(48\beta^2-R)}\, {\rm
arctanh}\Big(\fft{\sqrt{48\beta^2-R}}{4\sqrt3\,
\beta}\Big)\Big]\,.\label{d4prelag2}
\end{equation}
This somewhat more complicated theory is convergent at
$R_0=48\beta^2$, namely
\begin{equation}
F(R_0)=24\beta \sigma_2\,,\qquad R_0 f''(R_0)=16 \beta
\sigma_2\,.\label{ghostfreecon1}
\end{equation}
Thus in this case, the equation (\ref{R0eom}) is satisfied.  As we
shall discuss in section 6, this theory is ghost free in this de
Sitter vacuum. On the other hand, the theory becomes singular at
$R=0$, since $f(R)\sim R\log R$ when $R\rightarrow 0$. Thus, $F$ is
divergent at $R=0$ in this case.

   Thus we find an interesting phenomenon in $f(R)$ gravities.  The
same cosmology (\ref{d4premetric}) can be generated by two different
classical actions, but neither theory can smoothly describe the full
evolution. In the perturbative flat region, the Lagrangian
(\ref{d4prelag}) is a better theory. The later inflationary epoch is
better studied by the theory (\ref{d4prelag2}). This is similar to
the phenomenon in differential geometry that a manifold typically
requires multiple different but overlapping coordinate patches to
cover it. More detailed analysis will be given in section 6.

\subsection{Killing spinor equations}

    In the previous subsection, we present two $f(R)$ gravities that
give rise to the same cosmological solution with varying $R$.  Such
an exact solution is possible because the two theories are special
in that they admit Killing spinor equations.  In this subsection, we
derive the condition on $f$ so that the $f(R)$ theories admit
Killing spinor equations.

  It is clear that for a generic function $f$ there can be no
consistent Killing spinor equations.  However, for some classes of
functions, the theories can admit Killing spinor equations. The
simplest example is that $f=R - (D-2) \Lambda_0$, whose Killing
spinor equation is well established and given by
\begin{equation}
\hat D_\mu\epsilon\equiv \Big(D_\mu +
\ft12\sqrt{\ft{-\Lambda_0}{D-1}}\,\Gamma_\mu\Big)\epsilon=0
\,,\label{ks0}
\end{equation}
where $D_\mu$ is a covariant derivative on a spinor, defined by
\begin{equation}
D_\mu\epsilon = \partial_\mu \epsilon + \ft14
\omega^{ab}_\mu{}\Gamma_{ab}\epsilon\,.
\end{equation}
Whilst the equation $D_\mu\epsilon=0$ has a clear geometric
interpretation that $\epsilon$ is a covariant constant spinor, the
extra $\Gamma_\mu$ term in (\ref{ks0}) lacks an immediate
explanation. One defining property for the Killing spinor equation
is that the $\Gamma$-matrix projected integrability condition gives
rise to the Einstein equations of motion, namely
\begin{equation}
0=\Gamma^\mu [\hat D_\mu, \hat D_\nu] \epsilon= \ft12 \Gamma^\mu
(R_{\mu\nu} - \Lambda_0 g_{\mu\nu}) \epsilon\,.\label{int0}
\end{equation}
Thus we see that the projected integrability condition is satisfied
provided that the metric is Einstein with cosmological constant
$\Lambda_0$. It follows that without the extra $\Gamma_\mu$ term in
(\ref{ks0}), the Killing spinor equation would become irrelevant to
the theory with the cosmological constant. Conversely, the
significance of the above projected integrability condition is the
following. Although the existence of Killing spinor (\ref{ks0}) for
a background does not in general imply that it always satisfies the
equations of motion $R_{\mu\nu}=\Lambda_0 g_{\mu\nu}$, if one of the
Killing vectors constructed from the Killing spinors is time-like,
the background then indeed satisfies. This provides a powerful tool
for constructing exact solutions, since first-order equations are
much more manageable than the second-order ones. (The integrability
condition without the $\Gamma$-matrix projection is related to
Riemann tensor, and hence not related directly to the Einstein
equations of motion. This implies that not all solutions have
Killing spinors, even if the theory admits the Killing spinor
equation.)

We are now in the position to  derive the condition on $f$ so that
the theory admits Killing spinor equations, whose $\Gamma$-matrix
projected integrability conditions are analogous to (\ref{int0}).
Such Killing spinor equations were constructed recently for
non-supersymmetric theories involving the metric, a dilaton and a
form field \cite{lpw,lw,llw}.  It turns out the condition is very
restrictive and the only known non-trivial examples are either the
low-energy effective action of the bosonic string up to the
$\alpha'$ order \cite{lpw,lw} or the Kaluza-Klein theory with a
non-trivial scalar potential \cite{llw}. In $f(R)$ gravity, we
propose
\begin{equation}
{\cal D}_\mu\epsilon\equiv \Big(D_\mu + W(F(R)) \Gamma_\mu\Big)
\epsilon=0\,,\qquad \Big(\Gamma^\mu \nabla_\mu F +
U(F(R))\Big)\epsilon= 0\,,\label{ks}
\end{equation}
where the functions $W$ and $U$ are to be determined. The first
equation is the natural generalization of (\ref{ks0}).  The second
equation is inspired by the fact that $f(R)$ gravities are
effectively a special, {\it albeit} inconvenient,  class of
Brans-Dicke theories.  In supergravities, the left-hand sides of
equations in (\ref{ks}) would be the supersymmetric variations for
the gravitino and dilatino fields, and $F, W$ are analogous to the
scalar and the superpotential for the scalar.  If $f$ is linear in
$R$ and hence $F$ is a constant, we must set $U=0$ and we recover
the previous example.

To establish the relevance of these two Killing spinor equations
(\ref{ks}) to $f(R)$ gravities, we follow the procedure developed in
\cite{lpw,lw,llw}.  We first act on the second equation with
$\Gamma^\nu \nabla_\nu$, which gives
\begin{equation}
(\Box F - U(\dot U + 2 (D-1) W))\epsilon=0\,.\label{int1}
\end{equation}
In this paper, a dot is always denoted as a derivative with respect
to $F$.  Note that for any function $X$, we have $\dot X=X'/f''$, by
the virtue of the Leibnitz' chain rule.  The $\Gamma$-matrix
projected integrability condition for the first equation in
(\ref{ks}) is somewhat more involved, and we find it is given by
\begin{equation}
\Gamma^\mu [{\cal D}_\mu,{\cal D}_\nu]\epsilon = \ft12\Gamma^\mu
{\cal R}_{\mu\nu}\epsilon + \ft12 \Gamma^\mu X_{\mu\nu}\,,
\end{equation}
where ${\cal R}_{\mu\nu}$ is given by (\ref{Rmunueom}) and
$X_{\mu\nu}$ is given by
\begin{eqnarray}
X_{\mu\nu} &=& \Big(
2\dot W U + 4(D-1) W^2 + \fft{4(D-2)W \dot W U}{2W-\dot U}
-\fft{1}{2(D-1)F} ( f-2R F)\Big)g_{\mu\nu}\cr %%
&& +\Big(\fft{1}{F} - \fft{2(D-2)\dot W}{2W-\dot
U}\Big)\nabla_\mu\nabla_\nu F\,.
\end{eqnarray}
Thus for an $f(R)$ theory to admit consistent Killing spinor
equations (\ref{ks}), we must have
\begin{eqnarray}
&&F=\fft{2W-\dot U}{2(D-2)\dot W}\,,\qquad U(\dot U + 2 (D-1)W) =
\fft{1}{2(D-1)} (D f - 2 R F)\,,\cr &&2\dot W U + 4(D-1) W^2 +
\fft{2U W}{F} -\fft{1}{2(D-1)F} ( f-2R F)=0\,.\label{kscon0}
\end{eqnarray}
These equations can be reduced, giving rise to
\begin{eqnarray}
U=-\fft{4D(D-1) W^2 + R}{4(D-1)\dot W}\,,\qquad \dot W
=\fft{(4D(D-1)W^2 + R)W}{(4(D-1)(D-2) W^2+R) F - f}\,.\label{kscon1}
\end{eqnarray}
Thus, once $W$ and $f$ are known, $U$ comes out straightforwardly.
It is advantageous to express all functions in terms of the variable
$R$, in which case, the second equation above becomes
\begin{equation}
f'' -\fft{\Big(4(D-1)(D-2) W^2+ R\Big)
W'}{\Big(4D(D-1)W^2+R\Big)W}\,f' +
\fft{W'}{\Big(4D(D-1)W^2+R\Big)W}\,f=0\,.\label{feom}
\end{equation}
One way to view this equation is that it is a non-linear
differential equation of $W$ for a given $f$.  Owing to the
non-linearity, however, a solution for $W$ does not always exist for
a generic function $f$. An alternatively view point is that
(\ref{feom}) is a second-order linear differential equation for $f$.
Thus (\ref{feom}) must have solutions of $f$ for any given $W$, even
though the analytical expression for $f$ may not exist. This is very
much parallel to the gravity/scalar system, where a generic scalar
potential may not admit a superpotential, but any superpotential can
yield a potential.  The difference is that here there is no simple
expression directly for $f$ in terms of function $W$, but instead
$f$ has to be solved {\it via} the second-order linear differential
equation. What is curious is that for a given $W$, there can exist a
two-parameter (one non-trivial) family of $f(R)$ theories, since
second-order linear differential equations tend to give two
solutions associated with two integration constants.  In other
words, there are two different $f(R)$ gravities for one $W$. Of
course the function $U$ is also different. This is different from
the usual gravity/scalar system whose scalar potential can be
expressed simply in terms of a superpotential when the theory admits
Killing spinor equations.  In section 2.2, we have demonstrated such
two different $f(R)$ gravities can give rise to the same solution.
Furthermore, the cosmology at two different evolution epochs is
better studied by either one or the other $f(R)$ theories.

To summarize, we give a simple way of constructing $f(R)$ gravity
with Killing spinor equations. We can begin with a function $W$,
from which the $f(R)$ is determined by (\ref{feom}), and the
expression for $U$ follows straightforwardly from (\ref{kscon1}).
The corresponding Killing spinor equations are then (\ref{ks}).

  It should be emphasized that the Killing spinor equations we have
discussed are for the theory with generic backgrounds rather than
for a specific solution.  For an $f(R)$ theory with such Killing
spinor equations, we can check whether a specific solution preserves
Killing spinors and obtain the fraction of the maximally allowed
Killing spinors that survive.  For example, we can establish whether
the (A)dS vacua satisfying (\ref{R0eom}) admit any Killing spinor
that satisfies (\ref{ks}). Furthermore, armed with these Killing
spinor equations, we can construct new solutions. As mentioned
earlier, the Killing spinor equations (\ref{ks}) can be viewed as
the first integrals of the Einstein equations of motion. For a
background with a Killing spinor $\epsilon$ satisfying (\ref{ks}),
the integrability conditions imply, as we have shown,
\begin{equation}
{\cal R}\epsilon =0 \,,\qquad \Gamma^\mu {\cal
R}_{\mu\nu}\epsilon=0\,,\label{ksintsum}
\end{equation}
where ${\cal R}$ and ${\cal R}_{\mu\nu}$ are defined in
(\ref{traceeom}) and (\ref{Rmunueom}) respectively. Thus we see
immediately that the trace equation (\ref{traceeom}) is
automatically satisfied, and the second equation in (\ref{ksintsum})
takes the same form as (\ref{int0}). Note that the first equation
above is implied by the second equation, and hence it is not
independent. If the Killing vector constructed from the Killing
spinor $\epsilon$, namely $K^\mu(\epsilon) = \bar \epsilon
\Gamma^\mu \epsilon$, is {\it e.g.}~time-like, it is straightforward
to demonstrate that the equations of motion (\ref{Rmunueom}) are
satisfied as well. Thus for our $f(R)$ gravity with Killing spinor
equations, we can obtain a class of solutions by solving only the
Killing spinor equations (\ref{ks}) with the corresponding time-like
Killing vector, instead of solving the more difficult Einstein
equations directly. This class of solutions are analogous to the BPS
solutions in supergravities, and we shall refer them as ``BPS'' even
though our $f(R)$ gravities are not supersymmetric.  Note that in
practice, our task may not be about constructing the most general
``BPS'' solutions, but only some special solutions with some simple
ansatz. In this case, the existence of some Killing spinors that
satisfy the Killing spinor equations can determine the ansatz fully.
We can then simply substitute the result into (\ref{Gmunueom}) to
verify whether it is a solution or not.

\subsection{A few examples}
\label{examples}

Here we give some simple explicit examples of $f(R)$ gravities that
admit Killing spinor equations (\ref{ks}) and give the corresponding
$W$ and $U$ functions.  An obvious search is to consider quadratic
$f(R)$ gravity in $D$ dimensions, namely
\begin{equation}
f(R)=\sigma R - (D-2) \Lambda_0 + \alpha R^2\,,\label{quad}
\end{equation}
where $\sigma, \alpha$ and $\Lambda_0$ are constants.  For these
general parameters, we find that the Killing spinor equations do not
exist.  However, if the three parameters satisfy the following
constraint
\begin{equation}
\alpha \Lambda_0=-\fft{4(D-1)^2\sigma^2}{(D-2)(5D-2)^2}\,,
\label{quadcons}
\end{equation}
we find that they do exist, with
\begin{eqnarray}
U&=&-\fft{2c}{5D-2}\Big(D(D+2)\sigma + \ft13(D-4)(5D-2) F\Big)
F^{\fft12}\,,\cr %%
W&=&-\fft{c}{5D-2}\Big((D+2)\sigma - (5D-2)F \Big)F^{-\fft12}\,,\cr
c^2&=&-\fft{3}{16(D+2)(D-1) \alpha}\,.\label{quadsuppot}
\end{eqnarray}
Another example is also a $D$-dimensional theory:
\begin{equation}
f(R)=\sigma R\, (R + \beta)^{\fft{D-2}D}\,.\label{frexample2}
\end{equation}
We find that the corresponding $W$ and $U$ are given by
\begin{equation}
W=\fft{R}{2\sqrt{D(D-1)\beta}}\,,\qquad U=\fft{(D-2)R\Big((D-1)R +
\beta D\Big)}{D\sqrt{D(D-1)\beta}\,(R+\beta)^{2/D}}\,.
\end{equation}

In section 2.2, we have shown that the two four-dimensional $f(R)$
gravities (\ref{d4prelag}) and (\ref{d4prelag2}) give rise to the
same cosmological solution (\ref{d4premetric}). Both theories admit
Killing spinor equations, with the same $W$, {\it i.e.}
\begin{equation}
W=\fft{R}{48\beta}\,.
\end{equation}
However, these two theories have different $U$ functions.  For the
theory (\ref{d4prelag}), the function $U$ is
\begin{equation}
U=\fft{\sigma_1 R (R - 64\beta^2)}{16\beta \sqrt{48\beta^2-R}}\,.
\end{equation}
For the theory (\ref{d4prelag2}), we have
\begin{equation}
U=\fft{\sigma_2 \Big(12 \beta (32\beta^2 - R) \sqrt{48\beta^2-R} +
\sqrt3\, R (64\beta^2-R) {\rm
arccoth}\Big(\fft{4\sqrt3\,\beta}{\sqrt{48\beta^2-R}}\Big)\Big)
}{16\beta\sqrt{48\beta^2-R}}\,.
\end{equation}
More examples of $f(R)$ gravities that admit Killing spinor
equations will be given in subsequent sections.

\section{``BPS'' domain wall solutions}

\subsection{General properties}

As discussed in section 2, Einstein metrics arise naturally in
$f(R)$ gravities. However, for a generic $f(R)$ theory, any exact
solution beyond constant $R$ is more or less impossible to
construct, since one has to handle in general fourth-order
non-linear differential equations.  However, for our $f(R)$
gravities with Killing spinor equations, exact solutions with
varying $R$ can be obtained. In particular, we shall consider static
domain wall solutions, with the ansatz
\begin{equation}
ds_D^2 = dr^2 + e^{2A(r)} dx^\mu dx_\mu\,.\label{domainwallansatz}
\end{equation}
Making a natural choice of vielbein, $e^{r}=dr$, $e^i=e^A dx^i$, we
find that the non-vanishing spin connection is $\omega^{i}{}_r =
A_{,r} e^i$. The Ricci tensor and scalar are given by
\begin{eqnarray}
R_{rr}&=&-(D-1)(A_{,rr} + A_{,r}^2)\,,\qquad R_{\mu\nu} = -(A_{,rr}
+ (D-1)A_{,r}^2) g_{\mu\nu}\,,\cr %%
R&=&-2(D-1) A_{,rr} -D(D-1)A_{,r}^2 \,.\label{curvature}
\end{eqnarray}
For the metric ansatz (\ref{domainwallansatz}), the Killing spinor
equations (\ref{ks}) can be easily solved.  The Killing spinors are
given by
\begin{equation}
\epsilon = e^{\fft12 A} \epsilon_0\,,\qquad \Gamma_r
\epsilon_0=\epsilon_0\,,\label{ksres}
\end{equation}
where $\epsilon_0$ is a constant spinor in the $(D-1)$-dimensional
world-volume. The full set of equations of motion is now reduced to
simply
\begin{equation}
A_{,r}=-2W\,.\label{foeom}
\end{equation}
This is a tremendous simplification of the Einstein equation
(\ref{Gmunueom}). We see that the structure of domain wall is
completely determined by the function $W$.  This is analogous to
supergravities, where the BPS domain walls are solely determined by
the superpotential.

From the Killing spinors (\ref{ksres}), we can construct the Killing
vectors, namely $K^M(\epsilon_0)=e^{A} \bar\epsilon_0 \Gamma^M
\epsilon_0$, and hence $K^r=e^{A} \bar \epsilon_0\epsilon_0$,
$K^t=e^{A} \bar\epsilon_0\Gamma^t\epsilon_0$ and
$K^i=e^{A}\bar\epsilon_0 \Gamma^i\epsilon_0$.  Whether this Killing
vector can be time-like or not depends on dimensions and the
$\Gamma$-matrix properties. Since $\epsilon_0$ can be any constant
spinor in the $D-1$ dimensional spacetime $dx^\mu dx_\mu$, we would
expect that we could choose some appropriate $\epsilon_0$ such that
$K^r=0=K^i$, but $K^t\ne 0$.  In the case when $D$ is odd, and hence
$(D-1)$ is even, such an $\epsilon_0$ can be easily found. For
example, let us consider the convention that
$\bar\epsilon=\epsilon^\dagger \Gamma^t$ with Hermitian $\Gamma^i$
and anti-Hermitian $\Gamma^t$, the Killing spinors $\epsilon_0$
satisfying the above conditions are given by
\begin{equation}
\epsilon_0=(1+\gamma) \eta_0\,,
\end{equation}
where $\gamma=a\prod_i \Gamma^i$ with $a$ so chosen that
$\gamma^2=1$. It is straightforward to verify that the corresponding
$K$ is a time-like Killing vector. Following the discussion in the
previous section, the equation (\ref{foeom}) must satisfy
(\ref{Gmunueom}).  In even $D$ dimensions, the discussion is
somewhat more complicated. Let us consider two constant spinors
$\epsilon^\pm_0 =(1\pm \Gamma_1)\eta_0$, and denote $K^\pm$ as the
corresponding Killing vectors.  It is clear that $K^\pm$ are both
null vectors and hence we can choose a convention such that
$K=K^++K^-$ is time-like. Since we must have ${\cal R}_{\mu\nu}
K^\nu=0$, it is then straightforward to show that ${\cal
R}_{\mu\nu}=0$.  Note that this demonstration works in odd $D$
dimensions as well.

In fact, for such a simple background, it is quite easy to
demonstrate that (\ref{foeom}) indeed satisfies all the equations of
motion by simply substituting (\ref{foeom}) directly into
(\ref{Gmunueom}). From the expression for the Ricci scalar in
(\ref{curvature}), we obtain the first-order equation
\begin{equation}
W_{,r}=W' R_{,r}=\fft{R + 4D(D-1) W^2}{4(D-1)}\,.\label{Weom}
\end{equation}
Together with (\ref{feom}), we find that
\begin{equation}
F_{,r}=\fft{(R + 4(D-1)(D-2) W^2)F -f}{4(D-1)V}\,.
\end{equation}
It is now straightforward to establish that the full set of
equations of motion are all satisfied.

Now let us consider the general properties of the solution.  We use
$X$ to denote the right-hand side of (\ref{Weom}), namely
\begin{equation}
X(W)=\fft{R + 4D(D-1) W^2}{4(D-1)}\,.\label{xw}
\end{equation}
Here we treat $R$ as a function of $W$.  The equation (\ref{Weom})
is of the first order, and can be solved as
\begin{equation}
r-r_0=\int \fft{dW}{X(W)}\,.
\end{equation}
If $X(W)$ has a zero, such that
\begin{equation}
X(W) = X'(W_0) (W-W_0) + \cdots\,,
\end{equation}
we find that near the region of $W_0$, the solution is given by
\begin{equation}
W=W_0 + e^{X'(W_0)\, r}\,, \qquad
e^{A} = \exp\Big(-W_0\,r -\fft{e^{X'(W_0)\,r}}{X'(W_0)}\Big)\,.
\end{equation}
Thus if $X'(W_0)>0$, the metric becomes AdS when $r\rightarrow
-\infty$. The resulting metric is AdS horizon if $W_0<0$, and it is
AdS asymptotic boundary if $W_0>0$.  If on the other hand we have
$X'(W_0)<0$, the metric becomes AdS when $r\rightarrow +\infty$. The
resulting metric is AdS horizon if $W_0>0$, and it is AdS asymptotic
boundary if $W_0<0$.  Thus we see that near the region of $W=W_0$,
the solution is regular, approaching either the AdS horizon or the
AdS boundary.  Thus, If $X(W)$ has at least two roots, we can expect
smooth solutions that run from one AdS to the other, associated with
two adjacent roots.

\subsection{A class of exact solutions}

As discussed above, in order to construct smooth solutions, it is
necessary that the function $X(W)$ has two roots. In this
subsection, we consider a class of $W$, which is given by
\begin{equation}
R=4D(D-1)\Big((a-1) W^2 + b W + c\Big)\,.\label{Wchoice}
\end{equation}
where $a,b,c$ are constants.  It follows that $X(W)$ is quadratic:
\begin{equation}
X(W)=D(a W^2 + b W + c)\,,
\end{equation}
which has two roots, given by
\begin{equation}
2W_\pm =\lambda_\pm \equiv \fft{b\pm \sqrt{\Delta}}{a}\,.
\end{equation}
We require that the discriminant $\Delta\equiv b^2-4ac > 0$ so that
the two roots are real.  The $W$ equation (\ref{Weom}) implies that
\begin{equation}
W=-\fft{1}{2a}\Big(b +
\sqrt{\Delta}\,\tanh(\ft12D\sqrt{\Delta}\,(r+r_0))\Big)\,.
\end{equation}
It follows from (\ref{foeom}) that we have an explicit solution
\begin{equation}
e^{2A} = \Big(e^{b D r}\cosh^2(\ft12D\sqrt{\Delta}\,(r-r_0))
\Big)^{\fft{2}{aD}}\,.
\end{equation}
The metric is smooth with $r$ running from $-\infty$ to $+\infty$.
In both limits, the metric approaches AdS, namely
\begin{equation}
e^{2A}\rightarrow e^{2\lambda_\pm r}\,,\qquad \hbox{for}\qquad
r\rightarrow \pm \infty\,.
\end{equation}
The resulting effective cosmological constants in the AdS limits are
$\Lambda_\pm =-(D-1)\lambda_\pm^2$.

For the choice of $W$ given by (\ref{Wchoice}), it follows from
(\ref{feom}) that $f$ can be determined by the following
differential equation
\begin{eqnarray}
&&D(aW^2 + b W + c)(2(a-1)W+b)W f_{,WW}\cr
&&-\Big(4(a-1)W((aD-1)W^2+cD) + b((5aD-4D-2)W^2+cD)+ b^2D
W\Big)f_{,W}\cr &&+D(2(a-1)W+b)^2 f=0\,.\label{fweom}
\end{eqnarray}
For generic parameters $(a,b,c)$, there is no analytical solution
for $f$; however, for some special choices of these constants, we
obtain explicit $f(R)$ gravities. One example is that $b=0$.  In
this case, the $f(R)$ is given by two hypergeometric functions
\begin{equation}
f=\sigma_1 W^3\,\,  {}_2F_1(x_-,x_+;-\fft12; -\fft{aW^2}{c}) +
\sigma_2\,\, _2F_1(y_-,y_+; \fft52; -\fft{aW^2}{c})\,,\label{fres}
\end{equation}
where $\sigma_1$ and $\sigma_2$ are integration constants properly
chosen so that the function is real, and
\begin{eqnarray}
x_\pm&=&\fft{2-3aD \pm \sqrt{4+ aD((a+8)D-12)}}{4aD}\,,\cr
y_\pm&=&\fft{2+3aD \pm \sqrt{4+ aD((a+8)D-12)}}{4aD}\,.
\end{eqnarray}
The expression for $W$ in terms of $R$ can be obtained from
(\ref{Wchoice}).  Another example is provided that $a=1$ and $c=0$,
in which case, we have
\begin{equation}
f(R)=\sigma R\, \Big(R +
4D(D-1)b^2\Big)^{\fft{D-2}D}\,.\label{frsimple1}
\end{equation}
Note that here we have not presented the other choice for $f$ which
involves hypergeometric functions.  This simple $f(R)$ theory were
presented earlier in section \ref{examples}.

\subsection{Randall-Sundrum II}

The Randall-Sundrum (RS) II scenario is characterized by the metric
profile of the type $e^{2A} = e^{-2k|r|}$, with positive constant
$k$ \cite{rs2}. In other words, the metric approaches the AdS
horizons at both $r\rightarrow \pm \infty$ limits, with one maximum
in the middle.  (Although $r$ is a non-compact coordinate, the
volume integration is finite with respect to $r$.) This can be
achieved in our domain wall solution by requiring $a<0$, $c<0$,
which ensures that
\begin{equation}
\lambda_+<0\,,\qquad \lambda_->0\,.
\end{equation}
Note that the constant $b$ can be arbitrary in this case. Our
solution describes in general the asymmetric RS II scenario when
$\lambda_+ +\lambda_-$ does not vanish, since the cosmological
constants $\Lambda_\pm$ are not equal at the two AdS horizons.  This of
course is not crucial for trapping gravity on the wall. A symmetric
RS II can be obtained by further requiring $b=0$, whose $f(R)$
theory is given by (\ref{fres}).

To study the trapping of gravity on the wall, it is advantageous to
express first the metric in the conformally-flat frame, namely
\begin{equation}
ds^2 = e^{2A(z)} (dx^\mu dx_{\mu} + dz^2)\,,
\end{equation}
where the coordinate $z$ is related to $r$ as follows
\begin{equation}
-\lambda_-\, z = e^{-\fft{b}{a} r} \Big(\fft{1 +
\exp(D\sqrt{\Delta}\, r)}{\cosh(\fft12D\sqrt{\Delta}\,
r)}\Big)^{\fft{2}{aD}}\, {}_2F_1( \fft{\sqrt{\Delta} - b}{a
D\sqrt{\Delta}}, \fft{2}{aD}; 1+ \fft{\sqrt{\Delta} - b}{a
D\sqrt{\Delta}}; -e^{D\sqrt{\Delta}\, r})\,.
\end{equation}
Note that we have set the inessential $r_0$ to zero. The linear
fluctuation of the graviton modes in the $(D-1)$-dimensional flat
world-volume in the context of $f(R)$ theory has not been studied
yet except for $D=5$ \cite{zly}.  Thus we shall focus our attention
on five dimensions.  Following the procedure outlined in \cite{zly},
we let $h_{\mu\nu}=e^{-3A/2} F^{-1/2} n_{\mu\nu} \phi(z)$, where
$n_{\mu\nu}$ is transverse and traceless, we find that
\begin{equation}
-\ft12 \phi_{,zz} + V\phi=0\,,
\end{equation}
where the Schr\"odinger potential $V=V_0+V_1$ contains two parts.
The first part is given by
\begin{equation}
V_0=\ft12 k^2 + \ft14(D-2) A_{,zz} + \ft18 (D-2)^2 A_{,z}^2\,,
\end{equation}
which is the same as that in \cite{rs2}. For our general domain wall
solution, we find
\begin{eqnarray}
V_0&=& \ft12 k^2+ \fft{D-2}{8a^2} \exp\Big[\ft{2}{a}(b r +
\ft{2}{D}\log\Big(\cosh(\ft12D\sqrt{\Delta}\,(r-r_0))\Big)
\Big]\times\cr %%
&&\Big(a D \Big({\rm sech}(\ft12D\sqrt{\Delta}\, r)\Big)^2 + \Big(b
+ \sqrt{\Delta} \tanh(\ft12D\sqrt{\Delta}\, r)\Big)\Big)\,.
\end{eqnarray}
The second part is the contribution from $F(R)$, and it was
established only for $D=5$ \cite{zly}.  It is given by
\begin{equation}
V_1=\fft{3A_{,z} F_{,z}}{4F} - \fft{F_{,z}^2}{8F^2} +
\fft{F_{,zz}}{4F}\,.
\end{equation}
Note that the conversion to the coordinate $z$ does not give a close
form for general parameters and one can appeal to the numerical
approach. For some choice of parameters, the explicit $V$ as a
function of $z$ can be obtained. For example, let $a=-2/D, b=0$ and
$c=D/2$, in which case, we have
\begin{equation}
e^{A}=\fft{1}{\cosh(Dr)}\,.
\end{equation}
It follows that $r=\fft{1}{D}{\rm arcsinh}(D z)$, and hence
\begin{equation}
V_0=\ft12 k^2 + \fft{D^2(D-2)(D^3 z^2 -2)}{8(D^2 z^2 +1)^2}\,.
\end{equation}
This potential profile is very much like the one obtained in
\cite{gre} and is capable of trapping gravity on the wall, but ours
is realized by $f(R)$ gravity, with
\begin{eqnarray}
f(R)&=&\sigma_1 |W|^3\,
{}_2F_1(\ft12(1-\sqrt{2-D}),\ft12(1+\sqrt{2-D}); \ft52;
\fft{4W^2}{D^2})\cr %%
&&+\sigma_2\, {}_2F_1(-1-\ft12\sqrt{2-D}, -1+\ft12\sqrt{2-D};
-\ft12; \fft{4W^2}{D^2})\,,\cr %%
W^2&=& \fft{2D^2(D-1)-R}{4(D-1)(D+2)}\,,\label{rslagex}
\end{eqnarray}
The alarming-looking complex arguments in the hypergeometric
functions do not prevent the $f(R)$ from being real provided that
$|W|\le D/2$.  Of course, in order to demonstrate the trapping of
gravity in the $f(R)$ theory, we also need to look at the
contribution from $V_1$. Let us consider the simpler example
associated with $\sigma_2$ in (\ref{rslagex}) in $D=5$.  We find
that
\begin{eqnarray}
V_1&=&\fft{25}{8(1+25z^2)^2}\Big[6 - 3\tanh^2\Big(\sqrt3\,{\rm
arcsin} \fft{5z}{\sqrt{1+25z^2}}\Big)\cr %%
&&\qquad-50z\sqrt{\fft{3+75z^2}{1+25z^2}}\, \tanh\Big(\sqrt3\,{\rm
arcsin} \fft{5z}{\sqrt{1+25z^2}}\Big)\Big]\,.
\end{eqnarray}
The Schr\"odinger potential $V=V_0+V_1$ becomes a bit more
complicated.  It has a local maximum $V=0$ at $z=0$, and two
negatives minimums when we increase $|z|$, and it becomes positive
until it hit a maximum before it approaches zero at $|z|=\infty$.
Thus we demonstrate that $f(R)$ gravity can easily reproduce the
smooth RS II scenario with thick domain walls. Note that we have
added in an absolute value symbol on $W^3$ so that the $f(R)$ is a
symmetric function of $W$.  This does not create a discontinuity in
(\ref{fweom}) at $W=0$ since it involves only up to the second-order
derivatives.  The Ricci scalar for these solutions runs from
$R_0=-D^3(D-1)$ on the AdS horizon at $r=-\infty$ to the maximum
value of $-2D^2(D-1)$, and then decreases and approaches $R_0$ again
on the AdS horizon at $r=+\infty$, and correspondingly $W$ runs from
$-D/2$ to $D/2$.  It is easy to verify that for (\ref{rslagex}), the
constant $R_0=-D^3(D-1)$ solution indeed satisfy the equation
(\ref{R0eom}). Furthermore, we have
\begin{equation}
F(R_0)= -\fft{3D \sinh(\ft12\sqrt{D-2}\,\pi)}{16(D+2)(D-1)
\sqrt{D-2}}\, \sigma_1
+\fft{\cosh(\ft12\sqrt{D-2}\,\pi)}{2D^2(D-1)}\,\sigma_2\,.
\end{equation}
The $f''(R_0)$ is divergent unless the constants $\sigma_1$ and
$\sigma_2$ are specifically related.  Since the expression becomes
very complicated, we shall only give the result for $D=5$.  In this
case, we have
\begin{equation}
\sigma_2=\fft{125\sqrt3}{56} \coth(\ft12\sqrt3\,\pi)\, \sigma_1\,,
\end{equation}
and
\begin{equation}
F(R_0)=\fft{5\sqrt3}{448}{\rm csch}(\ft12\sqrt3\,\pi)\,\sigma_1\,,
\qquad R_0 f''(R_0)=-\fft{75\sqrt3}{6272} {\rm
csch}(\ft12\sqrt3\,\pi)\, \sigma_1\,,\label{rs2check}
\end{equation}
where $R_0=-500$.

  More generally, the RS II scenario arises in any $X(W)$ given in
(\ref{xw}) with a profile that it has one positive, one negative
root and a maximum in between.

Note that an analytical domain-wall solution in quadratic $f(R)$
theory together with additional scalar with $\phi^4$ potential was
obtained in \cite{lzzl}.

\subsection{AdS wormholes}

The situation is quite different if we have
\begin{equation}
\lambda_+>0\,,\qquad \lambda_- <0\,.
\end{equation}
This can be achieved by requiring $a>0$ and $c<0$, while $b$ can be
arbitrary.  In this case, the solutions describe smooth AdS
wormholes that connect two AdS boundaries at $r\rightarrow \pm
\infty$, with no bulk singularity and horizon in between.  For
non-vanishing $b$, the wormhole connects two asymmetric AdS
boundaries with different $\Lambda_\pm$.  For $b=0$, the AdS
boundaries are symmetric. Let us present a relative simple example
with $a=2/D, b=0$ and $c=-D/2$. In this case the theory is given by
\begin{eqnarray}
f&=&\sigma_1 |W|^3
{}_2F_1(-\ft12(1+\sqrt{D-1}),-\ft12(1-\sqrt{D-1});-\ft12;-\ft{4}{D^2}
W^2)\cr %%
&&+\sigma_2\, {}_2F_1(1-\ft12\sqrt{D-1}, 1+\ft12\sqrt{D-1}; \ft52;
\ft{4}{D^2} W^2)\,,\cr %%
W^2&=& -\fft{2D^22(D-1)+ R}{4(D-1)(D-2)}\,.
\end{eqnarray}
We add an absolute sign in $W$ for the same reason explained in the
previous subsection. The solution is quite simple:
\begin{equation}
ds^2 = \cosh^2(Dr) dx^\mu dx_\mu + dr^2\,.\label{adswh}
\end{equation}
The metric reaches AdS boundaries with $R_0=-D^3(D-1)$. For $D=1 +
k^2$, the function $f$ is given by simple functions.  For example,
when $D=5$, we have
\begin{equation}
f(R)=\sigma_1 \sqrt{-(R + 200)^3} + \sigma_2 (R+50)\sqrt{R+ 500}\,.
\label{adstoads}
\end{equation}
For the $f(R)=\sigma_1 \sqrt{-(R + 200)^3}$ theory, we have
$F(R_0)=-15\sqrt{3}\sigma_1$, and hence the (\ref{R0eom}) is
satisfied. For the $f(R)=\sigma_2 (R+50)\sqrt{R+ 500}$ theory, on
the other hand, the $F(R_0)$ is divergent. Such a situation was
discussed in section 2.2.

It is worth commenting that although there is a no-go theorem in the
usual Einstein theory that the configuration with two AdS boundaries
connected in the bulk without a horizon separating them will violet
the energy condition \cite{nogoworm}. This no-go theorem can be
easily circumvented in higher-order derivative theories.  Smooth
wormholes with two AdS boundaries were constructed in Einstein
gravity with the Gauss-Bonnet term \cite{worm1,worm2, vazwormhole}.
Our examples demonstrate that wormholes arise naturally in $f(R)$
gravities as well.  More generally, wormhole solutions occur in any
$X(W)$, given in (\ref{xw}), with a profile that it has one positive
and one negative roots with a minimum in between.  Note that our
wormhole solutions are brane-like and static. Only stationary
brane-like wormholes were known to exist in Einstein gravity and
supergravities in higher dimensions
\cite{lmworm,blmpworm,lmw,lwworm}.

     We can also consider a different parametrization.  We set
$a=1$ and $c=0$, but with non-vanishing $b$.  In this case, one $f$
is given in (\ref{frsimple1}). The domain wall solution is given by
\begin{equation}
e^{2A} = \Big(1 + e^{D b (z-z_0)}\Big)^{\fft4{D}}\,.\label{flattoads}
\end{equation}
For $b>0$, this solution describes a wormhole that connects a flat
spacetime at $r\rightarrow -\infty$ to the AdS boundary at
$r\rightarrow +\infty$ with $R_0=-4D(D-1)b^2$.  Note that the
solution with $b<0$ is equivalent to $b>0$, by reversing the sign of
$r$. Similar solutions that connect the AdS boundary in one
asymptotic region to the flat spacetime in another have also been
found in supergravities \cite{blmpworm,lmw,lwworm}; however, these
solutions are stationary rather than static.

As discussed earlier, there is a different $f(R)$ theory that would
give rise to the same solution (\ref{flattoads}).  It is much more
complicated, given by
\begin{equation}
f=\sigma W(W+b) \Big(1 - \fft{2b}{(n-2)(W+b)}
+\Big(\fft{W}{W+b}\Big)^{\fft2D}\, {}_2F_1(-\ft2D, -\ft2D;
\ft{D-2}{D}; -\ft{b}{W})\Big)\,,
\end{equation}
where $W=R/(4D(D-1)b)$.  In the case of $D=4$, the expression is
simpler, given by
\begin{equation}
f(R)=\sigma R \Big[12 b - \sqrt{3(R+48b^2)}\, {\rm
arctanh}\Big(\fft{\sqrt{R+48b^2}}{4\sqrt3\,b}\Big)\Big]
\,.\label{flattoadsfr}
\end{equation}
This theory satisfies that $F(R_0)=24b\sigma$ and $R_0 f''(R_0)=16 b
\sigma$.

\subsection{RG flow from IR to UV}

In this case, we have $\lambda_+\lambda_->0$, which can be achieved
by requiring $b^2>4a c>0$. It is clear that $\lambda_\pm$ being both
positive is equivalent to the case with both being negative, by
merely reversing the sign of the coordinate $r$.  Let us thus
discuss the case with both being positive.  If $a>0$ and hence
$b>0$, we then have
\begin{equation}
\lambda_+>\lambda_->0\,.
\end{equation}
The metric describes a flow running from the AdS horizon with
$\Lambda_-$ at $r\rightarrow -\infty$ to the AdS boundary with
$\Lambda_+$ at $r\rightarrow +\infty$, which corresponds to the IR
and the UV regions in the dual conformal field theory respectively.
The cosmological constant $\Lambda_-$ in the IR region is smaller
than the $\Lambda_+$ in the UV region.  This type of behavior is
similar to the domain wall solutions in supergravities \cite{fgpw}.
Note that such solutions can be obtained in any $X(W)$ that has two
adjacent positive roots with a minimum in between. If instead, $a<0$
and hence $b<0$, the cosmological constant in the IR region is
bigger than that in the UV region.  Such a solution occurs in any
$X(W)$ that has two adjacent positive roots with a maximum in
between.

\subsection{On holographic $c$-theorems}

As we see in our explicit constructions of domain wall solutions,
$f(R)$ gravities are quite suitable for investigating the AdS/CFT
correspondence. One natural question is to examine the holographic
$c$-theorem.  One may view that $f(R)$ gravity is simply Einstein
gravity with an effective energy-momentum tensor built from the
Ricci scalar, with the equations of motion (\ref{Gmunueom})
expressed as
\begin{equation}
G_{\mu\nu} = T^{\rm eff}_{\mu\nu}\equiv G_{\mu\nu}-{\cal
G}_{\mu\nu}\,,\label{effTmunu}
\end{equation}
where $G_{\mu\nu} = R_{\mu\nu} - \ft12 g_{\mu\nu} R$ is the Einstein
tensor.  If one takes this point of view, one can follow the
procedure in \cite{fgpw} and define
\begin{equation}
a(r)\equiv \fft{\pi^{(D-1)/2}}{\Gamma(\fft12(D-1))A_{,r}^{D-2}}\,.
\end{equation}
Its variation with respect to the co-moving coordinate $r$ is given
by
\begin{equation}
a_{,r} = -
\fft{\pi^{(D-1)/2}}{\Gamma(\fft12(D-1))A_{,r}^{D-1}}\Big((T^{\rm
eff})^{t}{}_{t}- (T^{\rm eff})^r{}_r\Big)\,.
\end{equation}
The holographic $c$-theorem follows provided that $a_{,r}\ge0$,
which implies that the cosmological constant at the IR is smaller
than that in the UV. It is clear that whether the $c$-theorem holds
or not depends on the specific choice of $f(R)$ gravities, and we
have examples that both support and violet the $c$-theorem.

     A different approach is to treat $f(R)$ gravities as
Brans-Dicke theory, and the $c$-theorem is then dictated by the
corresponding scalar potential.  The third approach is treating
$f(R)$ theory as a pure gravity theory that can coupled to
additional matter so that the equations are now given by
\begin{equation}
{\cal G}_{\mu\nu} = T^{\rm mat}_{\mu\nu}\,.
\end{equation}
This follows the same approach of \cite{ms} where all ghost-free
curvature squared and cubic terms were considered. A monotonic
function $a$ can be found in these higher-order theories \cite{ms}.
It is of great interest to investigate the constraints on $f$ so
that the holographic $c$-theorem also holds and whether such
constraints are consistent with the conditions for Killing spinor
equations.

\section{``BPS'' cosmology}

\subsection{The set up}

     It is well-known that the de Sitter spacetimes also admit
Killing spinors, even though Einstein gravity with a positive
cosmological constant cannot be supersymmetrized.  The Killing
spinor equation is given by
\begin{equation}
\hat D_\mu\epsilon\equiv \Big(D_\mu + \ft{\rm
 i}2\sqrt{\ft{\Lambda_0}{D-1}}\,\Gamma_\mu\Big)\epsilon=0
\,,\label{dsks}
\end{equation}
This property of de Sitter space was exploited in constructing de
Sitter ``supergravities'' \cite{ds1,ds2} which are effectively the
analytical continuation of AdS supergravities. The function $W$ in
Killing spinor equations (\ref{ks}) is pure imaginary in this case.
We would like to assume implicitly that $W$ and $U$ in Killing
spinor equations are real. Thus for the purpose of studying
cosmology, we would like to rewrite the Killing spinor equations as
follows
\begin{equation}
{\cal D}_\mu\epsilon\equiv \Big(D_\mu + {\rm i}\, W \Gamma_\mu\Big)
\epsilon=0\,,\qquad \Big(\Gamma^\mu \nabla_\mu F + {\rm i}\,
U\Big)\epsilon= 0\,,\label{ks2}
\end{equation}
where $U$ is given by
\begin{equation}
U=\fft{R-4D(D-1) W^2}{4(D-1)\dot W}\,.
\end{equation}
It is important to note that our procedure of sending $W$ and $U$ to
imaginary values does not affect the reality of the function $f$,
which now satisfies
\begin{equation}
f'' -\fft{\Big(R- 4(D-1)(D-2) W^2\Big)
W'}{\Big(R-4D(D-1)W^2\Big)W}\,f' +
\fft{W'}{\Big(R-4D(D-1)W^2\Big)W}\,f=0\,.\label{feom2}
\end{equation}

      We now construct ``BPS'' cosmological solutions that admit Killing
spinors.  The ansatz is the FLRW metric with flat spatial directions
\begin{equation}
ds^2 = -dt^2 + a^2 dx^i dx^i\,.
\end{equation}
Requiring that the solution admit Killing spinors, the full set of
Einstein equations of motion is reduced to
\begin{equation}
\fft{a_{,t}}{a}=2W\,,\label{cosfo}
\end{equation}
which implies the following first-order equation
\begin{equation}
W_{,t} = Y(W)\equiv \fft{R-4D(D-1) W^2}{4(D-1)}\,.\label{cosfo1}
\end{equation}
We verify that (\ref{cosfo}) indeed satisfies (\ref{Gmunueom}).
Smooth cosmology emerges when $Y(W)$ has two adjacent roots
corresponding to two de Sitter spaces.  The cosmological evolution
runs from one de Sitter to the other.

    As an illustrative example, let us consider
\begin{equation}
R=4D(D-1) \Big((\alpha+1) W^2 + \beta W +
\gamma\Big)\,,\label{dsRchoice}
\end{equation}
such that $Y(W)=D(\alpha W^2 + \beta W + \gamma)$. The corresponding
$f(R)$ can be determined by the following second-order linear
differential equation
\begin{eqnarray}
&&\fft{4(\alpha+1)(\alpha D+1)W^3 + \beta(5\alpha D+4D+2) W^2+
D(4(\alpha+1)\gamma + \beta^2) W + \beta D\gamma}{D
W(2(\alpha+1)W+\beta)(\alpha W^2 + \beta W + \gamma)}f_{,W}\cr %%
&&\qquad+f_{,WW} + \fft{2(\alpha+1) W + \beta}{W(\alpha W^2 + \beta
W + \gamma)} f=0\,.
\end{eqnarray}
When $\beta =0$, the equation can be solved explicitly, giving
\begin{equation}
f=\sigma_1 W^3\,\,  {}_2F_1(\tilde x_-,\tilde x_+; \fft52;
-\fft{\alpha W^2}{\gamma}) + \sigma_2\,\, _2F_1(\tilde y_-,\tilde
y_+; -\ft12; -\fft{\alpha W^2}{\gamma})\,.\label{fresds}
\end{equation}
where $\sigma_1$ and $\sigma_2$ are integration constants properly
chosen so that the function is real, and
\begin{eqnarray}
\tilde x_\pm&=&\fft{3\alpha D-2 \pm \sqrt{4+ \alpha
D((\alpha-8)D+12)}}{4\alpha D}\,,\cr %%
\tilde y_\pm&=&\fft{-3\alpha D-2 \pm \sqrt{4+ \alpha
D((\alpha-8)D+12)}}{4\alpha D}\,.
\end{eqnarray}

The general ``BPS'' cosmological solution for (\ref{dsRchoice}) is
given by
\begin{equation}
a=\Big(e^{\beta D\, t}\cosh^2(\ft12D\sqrt{\Delta}\,
(t-t_0))\Big)^{-\fft{1}{\alpha D}}\,,
\end{equation}
where $\Delta=\beta^2-4\alpha\gamma >0$. The solution approaches
de Sitter spaces in both $t\rightarrow \pm \infty$ limits, with
$a\sim e^{\lambda_\pm t}$, where
\begin{equation}
\lambda_\pm =-\fft{\beta\pm \sqrt{\Delta}}{\alpha}\,.
\end{equation}
The corresponding cosmological constants of the de Sitter spaces in
these limits are $\Lambda_\pm=(D-1)\lambda_\pm^2$.  Depending on the
sign and values of $\lambda_\pm$, various cosmological scenarios
emerge.

If we take a view that $f(R)$ gravity is simply Einstein gravity
with an effective energy-momentum tensor, as in (\ref{effTmunu}),
the effective energy density and pressure for our cosmology are
given by
\begin{eqnarray}
\rho^{\rm eff}&=& \fft{a_{,t}^2}{a^2} =\fft{\Big(\beta + \sqrt{\Delta}
\tanh(\ft12D\sqrt{\Delta}\, t)\Big)^2}{\alpha^2}\,,\cr %%
p^{\rm eff} &=& -\fft1{D-1} \Big(\rho +
(D-2)\fft{a_{,tt}}{a}\Big)\cr &=&\fft{1}{2(D-1)\alpha^2}\Big[(\alpha
D(D-2)+ 2(D-1))\Delta \Big({\rm sech}(\ft12D\sqrt{\Delta}\,
t)\Big)^2\cr && \qquad - 2(D-1) \Big(\beta^2 + \Delta + 2\beta
\sqrt{\Delta}\, \tanh(\ft12D\sqrt{\Delta}\,t)\Big)\Big]\,.
\end{eqnarray}
Then we have
\begin{equation}
w^{\rm eff}(t)=\fft{p^{\rm eff}}{\rho^{\rm eff}} = -1 + \fft{\alpha
D(D-2)\Delta}{2(D-1)\Big(\beta \cosh(\ft12D\sqrt{\Delta}\, t) +
\sqrt{\Delta}\sinh(\ft12D\sqrt{\Delta}\, t)\Big)^2}\,.
\end{equation}
Note that in this discussion, we have set the inessential $t_0$ to
zero.  In the limits of $t\rightarrow \pm \infty$, we have $w^{\rm
eff}(\pm \infty)=-1$ as one would have expected.  Since $\Delta>0$,
the sign choice of $(w^{\rm eff} + 1)$ for the ``dark energy''
depends solely on the parameter $\alpha$.  An extremum occurs at
\begin{equation}
t=-\fft{2}{D\sqrt{\Delta}} {\rm arctanh}
\Big(\fft{\sqrt{\Delta}}{\beta}\Big)\,,
\end{equation}
corresponding to
\begin{equation}
(w^{\rm eff} + 1)_{\rm extremum} = \fft{D(D-2)\Delta}{8(D-1)\gamma}\,.
\end{equation}

\subsection{From inflation to ever-expanding universe}

The parameters $\alpha,\beta$ and $\gamma$ should be chosen such
that $\lambda_+$ and $\lambda_-$ are both positive. Furthermore we
must have $\lambda_+<\lambda_-$. In this model, the the universe
starts an inflation with a bigger cosmological constant $\Lambda_-$
at $t\rightarrow -\infty$ and end with an ever-expanding de Sitter
universe with a smaller cosmological constant $\Lambda_+$. Since we
have
\begin{equation}
\lambda_+ - \lambda_- = -\fft{2\sqrt{\Delta}}{\alpha}\,.
\end{equation}
It follows that we must have $\alpha >0$. This implies that for this
model the sign choice of $(w^{\rm eff} + 1)$ for the dark energy is
always positive throughout the evolution. Furthermore, $\beta$ must
be negative and $\gamma$ must be positive.  As a semi-realistic
model of our universe, we require
\begin{equation}
\fft{\Lambda_+}{\Lambda_-} << 1\,.
\end{equation}
This can be achieved by requiring $4\alpha\gamma/\beta^2<<1$, in
which case we have
\begin{equation}
\lambda_- \sim  -\fft{2\beta}{\alpha}\,,\qquad \lambda_+ \sim
-\fft{4\alpha\gamma}{\beta}\,.
\end{equation}
It can be shown that in the later part of the evolution, $f(R)\sim R
- R^2/\beta + {\cal O}(R^3)$.  Note that this type of semi-realistic
solutions emerge as long as $Y(W)$ has two positive roots with a
positive maximum in between.

    If instead we have $\lambda_+>\lambda_->0$, which can be
achieved by requiring $\alpha <0$, $\beta >0$ and $\gamma <0$, the
universe would start with a mild inflation, and inflates faster and
faster. Such a solution arises in general when $Y(W)$ has two
positive roots with a negative minimum in between.  Such a theory
provides a model for the multi-stage inflationary scenario.

\subsection{Bouncing universe}

    The universe bounces when $\lambda_-<0$, but $\lambda_+ >0$.
This occurs when $\alpha <0$ and $\gamma >0$.  The minimum $a$
occurs when $t=t_{\rm min}$, given by
\begin{equation}
t_{\rm min}-t_0=-\fft{2}{D\sqrt{\Delta}} {\rm
arctanh}(\fft{\beta}{\sqrt{\Delta}})\,.
\end{equation}
The minimum scale factor is given by
\begin{equation}
a_{\rm min}=\Big(1 -\fft{\beta^2}{\Delta}\Big)^{\fft{1}{\alpha
D}}\Big(\fft{\sqrt{\Delta}+\beta}{\sqrt{\Delta}-
\beta}\Big)^{\fft{\beta}{ \alpha D\sqrt{\Delta}}}\,.
\end{equation}
This type of bouncing universe emerges when $Y(W)$ has one positive
and one negative adjacent roots with a positive maximum in between.

\subsection{Pre-big bang model}

     In the special limit, namely $\gamma=0$ and $\alpha=-1$, the
solution is simple, given by
\begin{equation}
a=\Big(1 + e^{\beta D(t-t_0)}\Big)^{\fft2{D}}\,.
\end{equation}
In this case, one $f(R)$ gravity takes a simple form, {\it i.e.}
\begin{equation}
f(R)=R\Big(4D(D-1) \beta^2-R\Big)^{\fft{D-2}{D}}\,.
\end{equation}
We shall not present the other $f(R)$ gravity that gives rise to the
exact same solution. The solution connects the flat $R=0$ region to
the $R=4D(D-1)\beta^2$ de Sitter space.  It can be used to model the
singularity-free inflation scenarios.  In particular it predicts a
``pre-big bang'' flat universe which bursts into inflation by the
non-perturbative effect of the higher-order curvatures.  The
four-dimensional case was discussed in section 2.

\subsection{Smooth crunching universe}

If we have $\lambda_->0$, but $\lambda_+<0$, the universe starts
with an inflation, but end with a big crunch.  What is interesting
is that usually such model encounters a curvature singularity at the
crunch. But in our solution, the universe shrinks in the manner of a
de Sitter space, and hence there is no singularity.

\section{Relating to the Brans-Dicke theory}

It is well-known that $f(R)$ gravity can be cast into the form of
Brans-Dicke theory by the Legendre transformation.  To see this, one
starts with the Lagrangian
\begin{equation}
{\cal L}=\sqrt{-g}\Big(f(\chi) + f_{,\chi}(\chi)
(R-\chi)\Big)\,.\label{legrender}
\end{equation}
Variation with respect to $\chi$ gives rise to
\begin{equation}
f_{,\chi\chi}(R-\chi)=0\,.
\end{equation}
Thus provided that $f_{,\chi\chi}\ne 0$, it follows that $\chi=R$,
and hence (\ref{legrender}) is the usual $f(R)$ theory.
Alternatively, one can define
\begin{equation}
\varphi=f_{,\chi}(\chi)\,,\label{converstion}
\end{equation}
and hence the $f(R)$ gravity is equivalent to the Brans-Dicke theory
of the type
\begin{equation}
{\cal L}=\sqrt{-g}\Big(\varphi R + f(\chi(\varphi))- \varphi
\chi(\varphi)\Big)\,.
\end{equation}
This is a special class of Brans-Dicke theory with no manifest
kinetic term for $\varphi$.  The conversion of $f(R)$ gravity to the
Brans-Dicke theory requires finding the inverse function of $F=f'$,
which in general does not have explicit analytical form.  In most of
our examples that admit Killing spinor equations discussed in this
paper, the $f(R)$ theories are better discussed on their original
form, rather than converting to the corresponding Brans-Dicke
theories.  There are couple of examples we find that can be
converted into the gravity/scalar system, where the scalar
potentials are expressed in terms of simple functions.

\subsection{A quadratic $f(R)$ theory}

The first example is the quadratic $f(R)$ gravity given in section
(\ref{examples}). This is a particular simple example, since $F$ is
a linear function with a simple inverse.  Using the procedure above,
we find that the gravity/scalar theory in Einstein frame is given by
\begin{equation}
{\cal L}=\sqrt{-g} (R - \ft12(\partial \phi)^2 - V)\,,\label{scalar}
\end{equation}
where the scalar potential can be expressed in terms of a
superpotential, namely
\begin{equation}
V=\widetilde W_{,\phi}^2 - \fft{D-1}{2(D-2)} \widetilde W^2\,.
\end{equation}
We find that
\begin{eqnarray}
\widetilde W &=& \fft{4\sqrt2\,(D-1)c}{3(5D-2)} \Big(3(D+2)\sigma - (5D-2)
e^{a_1 \phi}\Big) e^{a_2\phi}\,,\cr %%
a_1 &=& \sqrt{\fft{D-2}{2(D-1)}}\,,\qquad a_2 =-
\fft{D}{2\sqrt{2(D-1)(D-2)}}\,.\label{tildew}
\end{eqnarray}

    The domain wall solution in the $f(R)$ gravity can be obtained
by treating $R$ as the coordinate, rather than $z$.  In other words, we
have
\begin{equation}
dz= \fft{dz}{dR} dR=\fft{W_{,R}}{X(R)}dR\,,
\end{equation}
where
\begin{equation}
X(R)=\fft{R + 4D(D-1) W(R)^2}{4(D-1)}\,.
\end{equation}
The function $A$ is now given by
\begin{equation}
A=-\int \fft{2W W_{,R}}{X(R)} dR\,.
\end{equation}
The domain wall approaches the AdS boundary at $R=R_0$ where
$X(R_0)=0$.

\subsection{Another example}

Another example is provided by (\ref{frexample2}).  In $D=4$, we
have
\begin{equation}
{\cal L}_4=R\sqrt{R+\beta}\,.
\end{equation}
The resulting scalar/gravity system (\ref{scalar}) has a complicate
scalar potential, given by
\begin{equation}
V=-\ft2{27} \Phi^{-2} \Big(\Phi^2 - 3\beta + \sqrt{\Phi^4 + 3\beta
\Phi^2}\Big)\Big ( -3 \Phi + \sqrt{2\Phi^2 + 3\beta + \sqrt{\Phi^4
+3 \beta \Phi^2}}\Big)\,,
\end{equation}
where $\Phi=e^{2\phi/\sqrt3}$.

    It should be pointed out that for the majority of our $f(R)$
gravities that admit Killing spinor equations, it is unnatural to
convert them to the Brans-Dicke theory.  If one insists on doing so,
the philosophy should be applied to Einstein gravity with a
Gauss-Bonnet term where the $R + \alpha R^2$ part should be
converted to the Brans-Dicke theory as well.  The consequence is
that Einstein gravity with Gauss-Bonnet term should be viewed as
the Brans-Dicke theory coupled with the Ricci and Riemann tensor square
terms. This formalism is clearly less elegant than the original pure
gravity formalism.

\section{Linear spectrum in (A)dS}

As was discussed in section 3, $f(R)$ gravity admits (A)dS metrics
as its vacuum solutions.  We have constructed a large number of
``BPS'' domain wall and cosmological solutions that run from one
(A)dS to the other.  It is a formidable task to examine the
stability of these solutions.  In this section, we study the linear
fluctuation of $f(R)$ gravity in such a (A)dS vacuum instead.  As
has been discussed in section 2, there are two types of (A)dS vacua
that could arise in $f(R)$ gravities.  The first type is the usual
one that satisfies (\ref{R0eom}).  The second type is the one we
discovered in this paper and it is characterized by the divergent
$F(R_0)$. In this section, we shall be only concerned with the
linearization $f(R)$ gravities around the (A)dS vacua of the first
type.  Linearized $f(R)$ gravity in such AdS$_4$ were studied in
\cite{myung}.  For the linear perturbation $g_{\mu\nu} \rightarrow
g_{\mu\nu} + h_{\mu\nu}$, we impose the gauge condition
\begin{equation}
\nabla^\mu h_{\mu\nu}=\nabla_\nu h\,.
\end{equation}
This gauge condition is different from the usual de Donder gauge,
but it is more effective to use in theories with a cosmological
constant since it implies the vanishing of the trace scalar mode in
Einstein gravity with a cosmological constant.  It has been adopted
in recent studies in critical gravities
\cite{lss,lpcritical,dllpst}.  We find that the linearized equation
of (\ref{Gmunueom}) becomes
\begin{eqnarray}
-\ft12 f'(R_0) \Big(\Box -\fft{2\Lambda}{D-1}\Big) H_{\mu\nu} &=&
0\,,\label{linearH}\\
-(D-1)\Lambda\, f''(R_0) \Big(\Box - m^2\Big) h
&=&0\,,\label{linearh}
\end{eqnarray}
where
\begin{eqnarray}
H_{\mu\nu} &\equiv& h_{\mu\nu} - \ft1{D} g_{\mu\nu} h -
\fft{2(D-1)f''(R_0)}{(D-2)f'(R_0)} J_{\mu\nu}\,,\cr J_{\mu\nu}
&\equiv& (\nabla_\mu\nabla_\nu - \ft1{D} \Box)h\,,\qquad m^2=-
\fft{R_0f''(R_0) -\ft12 (D-2) f'(R_0)}{(D-1)f''(R_0)}\,.
\label{lineareom}
\end{eqnarray}
To derive the above equations, we have made use of the following
formulae
\begin{equation}
[\Box, \nabla_\mu] h= \Lambda \nabla_\nu h\,,\qquad [\Box,
\nabla_\mu\nabla_\nu] h = \fft{2D\Lambda}{D-1} J_{\mu\nu}\,.
\end{equation}
It is clear that $H_{\mu\nu}$ is traceless; it is also transverse by
the virtue of the equation of motion for $h$. In the above, we
assume that $R_0, f(R_0), f'(R_0)$ and $f''(R_0)$ are all
non-vanishing.  Thus we see that in general, in addition to the
massless spin-2 graviton mode, there is also a massless scalar trace
mode. However, there is no higher-order propagator for both modes,
unlike the case in theories with more general higher curvature
invariants.  This is consistent with the fact that in terms of
physical degrees of freedom, $f(R)$ gravity is equivalent to a
special class of the Brans-Dicke theory.  The lacking of
higher-order propagators implies that there is no critical
phenomenon as those discussed in \cite{lss,lpcritical,dllpst}.

In the special case, where $f''(R_0)=0$, the equation
(\ref{linearh}) implies that $h=0$, and the theory contains only the
massless graviton, as in the case of Einstein gravity.  If
$f'(R_0)=0$, graviton $H_{\mu\nu}$ no longer has its kinetic term.
If $f'(R_0)=0=f''(R_0)$, the theory has no propagating mode at all.
For example, the theory
\begin{equation}
{\cal L}_4=\sqrt{-g} (R-R_0)^3
\end{equation}
satisfies the criteria.  What is interesting is that although such a
theory does not have any perturbative propagating degrees of freedom, it
nevertheless admits the (A)dS Schwarzschild black hole solution.  This
particular aspect of the theory is similar to three-dimensional
Einstein gravity with a cosmological constant.

For general case with non-vanishing $f'(R_0)$ and $f''(R_0)$, the
ghost-free conditions are
\begin{equation}
f'(R_0)>0\,,\qquad R_0 f''(R_0)> 0\,.\label{ghostfree}
\end{equation}
The tachyon-free Breitenlohner-Freedman (BF) condition in AdS is given by
\begin{equation}
m^2 \ge \fft{D-1}{4D} R_0\,,\label{tachyonfree}
\end{equation}

We now examine the stability of some of our $f(R)$ gravities in the
(A)dS vacua. The first example to consider is the quadratic
Ricci-scalar action (\ref{quad}).  We have demonstrated in section 5
that this is equivalent to (\ref{scalar}).  We choose a convention
that the AdS fixed point for the superpotential (\ref{tildew})
occurs at $\phi=0$, which implies that
\begin{equation}
\sigma=-\fft{(D-4)(5D-2)}{3D(D+2)}\,.
\end{equation}
Expanding the scalar potential $V$ around $\phi=0$, we find that
\begin{equation}
V=\fft{D-2}{D} R_0 +\ft12 M^2 \phi^2 + \cdots\,,
\end{equation}
where
\begin{equation}
\alpha R_0 = \fft{4(D-1)^2}{3D(D+2)}\,,\qquad
M^2=\fft{(D-4)(3D^2-4D+4)R_0}{16(D-1)^3}\,.\label{R0susy1}
\end{equation}
It follows from (\ref{quadsuppot}) that the reality condition
requires that $\alpha <0$, and hence the vacuum is AdS. Furthermore,
we have
\begin{equation}
M^2-\fft{D-1}{4D} R_0 = -\fft{(D^2+2)^2}{16D(D-1)^3} R_0>0\,,
\end{equation}
hence the BF bound is satisfied.  Thus we find that the
scalar/gravity theory is both tachyon and ghost free in the AdS
vacuum with $R=R_0$.  Now let us examine the corresponding $f(R)$
gravity.  For $D\ne 4$, there are two AdS vacua, namely
\begin{equation}
R=R_0\qquad \hbox{and}\qquad R=\tilde R_0\equiv
\fft{D-4}{3(D+2)\alpha}\,,
\end{equation}
where $R_0$ is given by (\ref{R0susy1}).  The $R=R_0$ vacuum is
``BPS'', whilst the $R=\tilde R_0$ one is not.  It is easy to verify
that $F(R_0)=1$ and $R_0 f''(R_0)=2\alpha R_0 > 0$. The $m^2$
calculated from (\ref{lineareom}) is exactly the same as $M^2$.  The
situation is quite different for the non-``BPS'' vacuum with
$R=\tilde R_0$. Although there is no tachyon, the spin-2 graviton is
a ghost field. If we reverse the overall sign of the action, the
spin-0 trace mode becomes a ghost. That the ``BPS'' vacuum is stable
whilst the non-``BPS'' vacuum is unstable is consistent with our
expectation.

The second example we would like to examine is the two theories
given in section 2.2.  Both theories (\ref{d4prelag}) and
(\ref{d4prelag2}) can give rise to the same cosmological solution
(\ref{d4premetric}) which describes an evolution from the flat
spacetime to the inflationary de Sitter vacuum.  It is clear that in
the flat region, (\ref{d4prelag}) is a good perturbative theory.
However, in the region where $R=R_0\equiv 48\beta^2$, the theory
becomes singular with divergent $F(R_0)$ and $f''(R_0)$.  On the
other hand, theory (\ref{d4prelag2}) is opposite.  In the $R=0$
region, the theory is singular, but it is well behaved in the
$R=R_0$ region, with $F(R_0)$ and $f''(R_0)$ given in
(\ref{ghostfreecon1}), and hence the vacuum is ghost free provided
that $\beta \sigma_2>0$.  Note that the same conclusion also holds
for the pair of theories (\ref{flattoadsfr}) and (\ref{frsimple1})
in $D=4$.  Furthermore the BF bound for the theory
(\ref{flattoadsfr}) in the ``BPS'' AdS vacuum is satisfied, and
hence the theory is both ghost and tachyon free.

Thus we see an interesting phenomenon in our $f(R)$ gravities, which
we have mentioned in section 2. For a giving Killing spinor
equation, and hence one ``BPS'' domain wall or cosmological
solution, there can be two $f(R)$ gravities.  For a solution that
connects two different AdS vacua with $\Lambda_+$ and $\Lambda_-$,
one $f(R)$ theory is well-defined in the $\Lambda_+$ vacuum with no
ghost and tachyon, but becomes singular at $\Lambda_-$, and vice
versa for the other $f(R)$.

The third example we consider is the five-dimensional $f(R)$ theory
(\ref{adstoads}) that admits the AdS wormhole solution
(\ref{adswh}). The AdS wormhole is symmetric with both AdS
boundaries having the same $R_0=-500$.  It is clear that the theory
associated with $\sigma_2$ is singular.  On other hand, the theory
associated with $\sigma_1$, namely
\begin{equation}
f(R)=\sigma_1\sqrt{-(R+200)^3}\,,
\end{equation}
is well defined in the vacuum.  We have $F(R_0)=15\sqrt3\,\sigma_1$
and $R_0f''(R_0)=25\sqrt3\,\sigma_1/2$, and hence the vacuum
fluctuation is ghost free for positive $\sigma_1$.  The mass square
of the spin-0 mode is given by $m^2=-100$, which precisely saturates
the BF bound. This theory is different from the second example, in
that both AdS boundaries have the same $R_0$ and hence one theory is
needed instead of having to have both theories to patch different
regions.

    The last example we shall examine is the smooth Randall-Sundrum
II solutions discussed in section 3. For the simpler case
(\ref{rslagex}) in $D=5$, both $F(R_0)$ and $R_0f''(R_0)$ are given
in (\ref{rs2check}). They cannot be both positive and hence the AdS
vacuum suffers from having a ghost field. (It is easy to obtain
$m^2=300$ for the scalar mode, and hence it is not a tachyon.) Of
course, this is only one example of many possible RS II solutions,
and it is of interest to investigate whether such a ghost problem of
the ``BPS'' RS II in $f(R)$ gravities is generic or not.
Furthermore, we have imposed that $f''(R_0)$ be finite, which is not
entirely clear to be necessary.

\section{Conclusion}

    In this paper, we follow the procedure outlined in
\cite{lpw,lw,llw} and obtain the condition on the subclass of $f(R)$
theories that admit Killing spinor equations.  We present many
examples of such $f(R)$ gravities.  One advantage of our theories is
that the Killing spinor equations reduce the fourth-order Einstein
equations for the domain wall and FLRW ansatze to very simple
first-order equations, and hence exact solutions can be constructed.

    For domain wall solutions, we find exact smooth examples that
describe the RS II scenario, AdS wormholes and the RG flow from the
IR to the UV. In all these solutions, the metric runs from one AdS
to another.  This is very different from other higher-derivative
theories such as Lovelock gravities with the Gauss-Bonnet term,
which also have multiple AdS vacua, but have no known flow running
from one to the other. Our examples demonstrate that $f(R)$ is a
fruitful arena to investigate and apply the AdS/CFT correspondence.

    Rich classes of exact and smooth cosmological solutions also
emerge in our $f(R)$ gravities.  We find a semi-realistic
cosmological solution that evolves from an inflationary starting
point to end with an ever-lasting expanding universe with a much
smaller cosmological constant. We also find a pre-big bang model
where a flat universe bursts into inflation by the non-perturbative
effect of the higher-order curvature terms. In addition, we find
smooth bouncing and crunching universes.

  Since the cosmological evolution in our $f(R)$ gravities is solely
governed by the equation (\ref{cosfo}), it is a matter of finding
the right profile of $W(R)$ in order to fit the observational data.
However, one technical drawback is that for a given $W$, the $f(R)$
is not determined directly, but ${\it via}$ a second-order linear
differential equation, which may not have a close-form solution.
Nevertheless we have obtained many explicit examples in this paper.
Classically, it can be argued that this is not essential since $W$
gives all the information.  At the quantum level, the exact form of
$f(R)$ is likely to become much more important.  It is of interest
to investigate whether it is possible to compute the quantum effect
on the information given by the $W$ alone.

   The full analysis of the stability of our ``BPS'' solutions is
beyond the scope of this paper. Instead, we investigate the
stability of the (A)dS vacua that these metrics connect to. We study
the linearized gravity around the (A)dS vacua. We adopt the gauge
that was used previously for studying critical gravities
\cite{lss,lpcritical}. In general, $f(R)$ gravity consists of one
massless spin-2 and one-massive spin-0 modes.  We also obtain the
condition for which the spin-0 mode decouples so that the spectrum
is identical to that of Einstein gravity.  More exotic situation can
arise where a theory has no propagating degree of freedom, yet it
admits the Schwarzschild (A)dS black hole as a solution, analogous
to Einstein gravity with a cosmological constant in three
dimensions. We obtain the conditions for $f(R)$ theories to be
absent from the ghost and tachyon fields, and give a detail analysis
for a few examples.

There is an intriguing phenomenon in our $f(R)$ gravities.  As we
have mentioned, for a giving $W$ in the Killing spinor equations,
and hence one ``BPS'' domain wall or cosmological solution, there
can be two $f(R)$ gravities.  For the solution that connects two
different AdS vacua with $\Lambda_+$ and $\Lambda_-$, We find
examples that one $f(R)$ theory is well-defined in the $\Lambda_+$
vacuum with no ghost and tachyon, but becomes singular at
$\Lambda_-$, and vice versa for the other $f(R)$.  This suggests
that there can exist multiple classical $f(R)$ gravities that give
the same full cosmological evolution; however, different stages of
the evolution may select different specific theories for the quantum
description.  This is similar to the common phenomenon in
differential geometry that a typical manifold requires multiple
different but overlapping coordinate patches in order to cover it.

It was shown in \cite{lpwpseudo,llwpseudo} that non-supersymmetric
theories that admit Killing spinor equations can be
pseudo-supersymmetrized by introducing pseudo fermionic partners. In
these theories, it can be shown that the Lagrangian is invariant
under the pseudo-supersymmetric transformation rules up to the
quadratic order in fermions.  This suggests that there should be
pseudo-supersymmetric versions of our $f(R)$ gravities.  It is of
great interest to construct such $f(R)$ pseudo-supergravities.

To conclude, our construction of $f(R)$ gravities that admit Killing
spinor equations allows us to find exact ``BPS'' domain wall and
FLRW cosmological solutions with varying Ricci scalar $R$.  The
significance of these solutions is that they explore the function
$f(R)$ in contrast to the previously known solutions with fixed $R$.
This opens a new door to study both the AdS/CFT correspondence and
cosmology in the context of $f(R)$ gravities.  Our construction is
based on $f(R)$ theories in the metric formalism, and hence it is
natural to extend our discussion to the Palatini formalism where
both the metric and the connection are assumed to be independent
variables. It is also of interest to investigate whether the Killing
spinor equations of our $f(R)$ gravities can be extended to include
matter.

\section*{Acknowledgement}

We are grateful to Kai-Nan Shao for useful discussions, and grateful
to KITPC, Beijing, for hospitality during the course of this work.
Liu is supported in part by the National Science Foundation of China
(10875103, 11135006) and National Basic Research Program of China
(2010CB833000). L\"u is supported in part by the NSFC grant
11175269.

\end{document}